# Output Voltage Response Improvement and Ripple Reduction Control for Input-parallel Output-parallel High-Power DC Supply

Jianhui Meng, *Member, IEEE*, Xiaolong Wu, Tairan Ye, Jingsen Yu, Likang Gu, Zili Zhang, Yang Li

***Abstract-*** **A three-phase isolated AC-DC-DC power supply is widely used in the industrial field due to its attractive features such as high-power density, modularity for easy expansion and electrical isolation. In high-power application scenarios, it can be realized by multiple AC-DC-DC modules with Input-Parallel Output-Parallel (IPOP) mode. However, it has the problems of slow output voltage response and large ripple in some special applications, such as electrophoresis and electroplating. This paper investigates an improved Adaptive Linear Active Disturbance Rejection Control (A-LADRC) with flexible adjustment capability of the bandwidth parameter value for the high-power DC supply to improve the output voltage response speed. To reduce the DC supply ripple, a control strategy is designed for a single module to adaptively adjust the duty cycle compensation according to the output feedback value. When multiple modules are connected in parallel, a Hierarchical Delay Current Sharing Control (HDCSC) strategy for centralized controllers is proposed to make the peaks and valleys of different modules offset each other. Finally, the proposed method is verified by designing a 42V/12000A high-power DC supply, and the results demonstrate that the proposed method is effective in improving the system output voltage response speed and reducing the voltage ripple, which has significant practical engineering application value.**

## I. INTRODUCTION

A three-phase isolated AC-DC-DC power supply, consisting of a front-end AC-DC uncontrolled rectifier for cost-saving and a downstream DC-DC phase-shifted full-bridge converter, is widely used in industrial high-power DC supplies owing to the advantages of high power density, easy expansion, and electrical isolation [1]-[3]. The characteristics of the downstream phase-shifted full-bridge converter are the main factors that determine the performance of the power supply. When high-power applications are required, multiple such AC-DC-DC power supplies can be implemented with Input-Parallel Output-Parallel (IPOP) mode. In some special applications, such as electrophoresis and electroplating, the output voltage of the power supply is required to have an extremely fast response due to instantaneous changes in load. Additionally, due to the influence of the front-end uncontrolled rectifier, the output voltage ripple is usually large, and configuring many filter capacitors to meet the design requirements will increase the cost of the power supply. Therefore, it is urgent to solve the problems encountered by high-power AC-DC-DC supplies from two aspects: improving the response speed of the output voltage and reducing the ripple.

In terms of improving the output response speed of the DC power supplies, a number of control methods for isolated DC-DC converters to obtain good dynamic output characteristics have been studied in recent years. As in [4], [5], controllers are designed based on sliding mode control to improve the system response speed. Sliding mode control has high stability and strong robustness, but its switching frequency is unstable, and the resulting loss and electromagnetic interference problems reduce its engineering applications. A fuzzy logic controller is presented in [6] to control the output voltage of the DC-DC converter in view of its nonlinear characteristics. When using this method, it is necessary to formulate appropriate fuzzy logic rules to achieve a better control effect. Then, a unit prediction horizon binary search-based nonlinear Model Predictive Control (MPC) for phase-shift full-bridge DC-DC converter is proposed in [7]. But this method is very computationally intensive. Therefore, for the downstream DC-DC converter, it is of great significance and value to investigate a control method with perfect dynamic response and suitable for industrial applications.

Active Disturbance Rejection Control (ADRC) is an improved strategy based on traditional Proportional-Integral-Derivative (PID) control and has good anti-disturbance performance. Linear Active Disturbance Rejection Control (LADRC) is proposed in [9], which linearizes the traditional ADRC, reduces the number of adjustable parameters and provides a parameter tuning method, and inherits the advantages of the original controller. To improve the dynamic tracking response speed and anti-interference ability, a controller with LADRC that compensates for the error of the total disturbance is proposed, and the stability of the improved first-order LADRC is proved by the Lyapunov stability theory [10]. The disturbance suppression and noise rejection of an ADRC controller for a Permanent Magnet Synchronous Motor (PMSM) are presented in [11], [12]. However, these methods are mainly proposed for speed control systems. And for a more complex nonlinear system, the control effect is not ideal for a dynamically changing controlled system when the ADRC strategy is limited by its fixed control parameters. The LADRC control method is very suitable for the application scenario in this study. Therefore, improving the flexible adjustment capability of LADRC, which is applied to DC-DC converters, is one of the main tasks of this paper.

In addition, the DC output voltage of the isolated DC-DC converter has a periodic six-time AC frequency fluctuation component which is produced by the front-end uncontrolled rectifier. The conventional Proportional-Integral (PI) controller is adjusted repeatedly during the pulsation cycle to control the duty cycle of the converter to oscillate back and forth, and the fluctuation is poorly suppressed. Several studies have been conducted in the literature to address this issue [13]-[21]. In [13], [14], the Notch Filter with Current Disturbance Feedforward Strategy (NF-CDFS) is proposed to minimize the input harmonic current of the AC-DC converter. And the Feedback Linearization with Voltage Disturbance Feedforward Strategy (FLVDFS) is designed to suppress the output voltage ripple of DC-DC converter based on a generalized state-space averaging model. The above methods achieve the purpose of reducing ripple by changing the control strategy or adding devices and changing the structure, but these methods are relatively complicated to implement and difficult to achieve in engineering applications. Considering the economy of industrial applications, designing a simple and easy-to-implement method to reduce the steady-state ripple of isolated DC-DC converter output voltage has considerable economic and engineering application value in improving power quality.

To solve the above problems in practical applications, this paper investigates an improved Adaptive Linear Active Disturbance Rejection Control (A-LADRC) method to improve the output response speed and proposes a control strategy to improve the output ripple in terms of both single module improvement and multiple-unit cooperation. The main contributions of the paper can be summarized as follows:

1) To improve the output voltage response speed of the high-power DC supply, an A-LADRC which can adjust the bandwidth parameter value flexibly according to the output real-time value is proposed. The presented strategy, which combines the output value with the key bandwidth parameter through a designed formula, can ensure a faster response speed of the output voltage in a wider range.

2) To reduce the DC supply ripple, a control strategy is designed for a single module to adaptively adjust the duty cycle compensation according to the output feedback value. When multiple modules are connected in parallel, a hierarchical delay current sharing control strategy for centralized controllers is proposed to make the peaks and valleys of different modules offset each other.

3) The performance of the proposed control is fully validated with a controller-level hardware-in-the-loop platform and a 42V/12000A IPOP high-power DC supply. And the effectiveness and industrial application value of the proposed method are fully verified from different aspects.

The main contents of other parts of this paper are as follows. The second section introduces the system topology and modeling of the high-power supply. The third section introduces the proposed A-LADRC strategy to improve the output voltage response speed. In the fourth section, the ripple reduction control is studied. Then, the effectiveness and superiority of the proposed control strategy are verified by experiment results in the fifth section. Finally, the sixth section summarizes the work of this paper.

## II. System Operating Principle and Modeling Analysis

### A. System topology structure

Multiple AC-DC-DC power supplies can be implemented with IPOP mode for high-power applications. The DC supply with a parallel structure usually has a certain redundancy design to achieve high safety. The topology of the system studied in this paper is shown in Fig. 1. The system consists of twelve isolated AC-DC-DC modules with IPOP mode. This flexible design approach reduces the current stress on each module and increases the system power density.

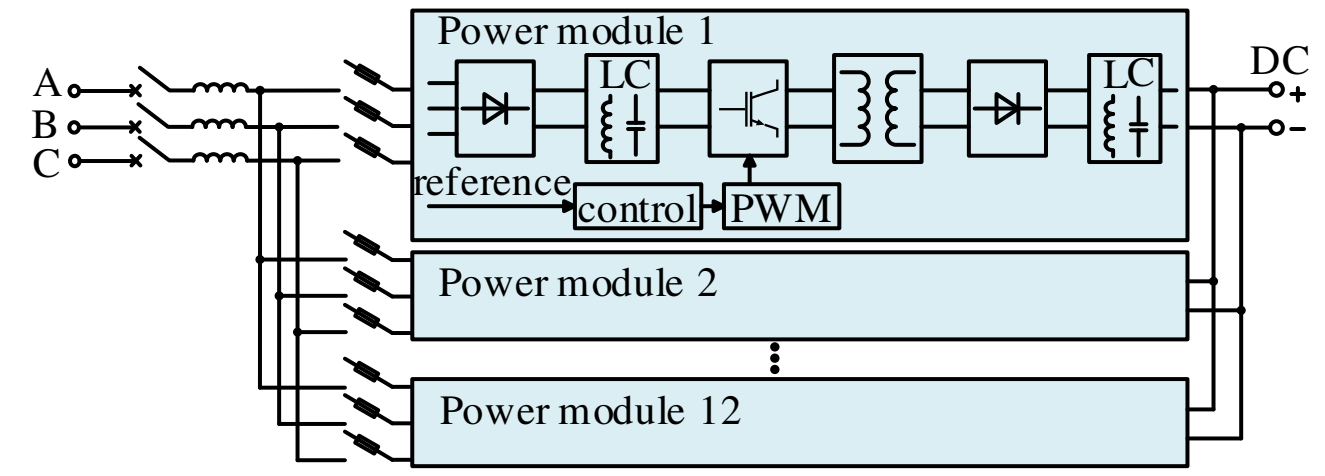

Fig. 1 Topology structure of the system

The topology structure of a single isolated AC-DC-DC module is shown in Fig. 2. The front-end of the module is a three-phase uncontrolled bridge rectifier. $u_i$ is the DC bus voltage obtained by the rectifier. $L_{lk}$ is the leakage inductance of the high-frequency transformer. $S_1$~$S_4$ are switches of the full bridge inverter. $L_2$ and $C_2$ are the output filter inductor and capacitor respectively. The output voltage $u_o$ is used to supply power to the load. $u_p$ is the primary voltage of the transformer and its duty cycle is $D$. $u_s$ is the secondary voltage and its duty cycle is $D_{eff}$; the transformer ratio is $n$:1. The downstream DC-DC high frequency converter adopts the phase-shifting Pulse Width Modulation(PWM) control.

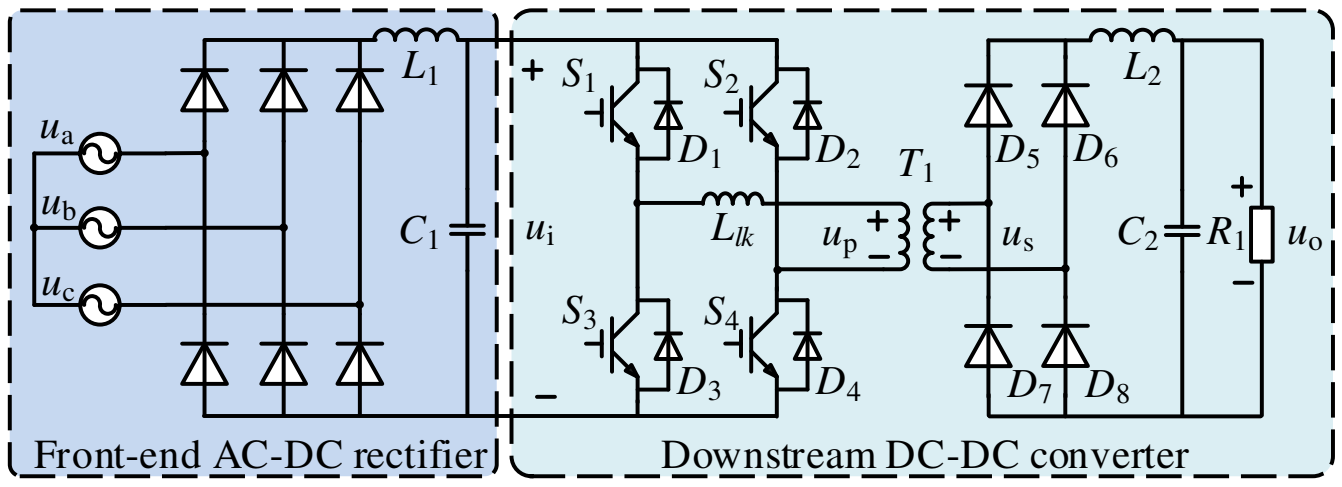

Fig. 2 Topology structure of the AC-DC-DC module

### B. Modeling and analysis of a single module

In the front-end three-phase bridge uncontrolled rectifier, the output voltage of the rectifier is a periodic non-sinusoidal function, in which the DC component includes harmonics of different frequencies [23]. The rectified output voltage can be obtained by the resolution of the Fourier series (take $m$ pulse waves as an example)

$$U_d = U_{d0} + \sum_{n=mk}^{\infty} b_n \cos n\omega t \tag{1}$$

where, $k$=1,2,3,…; $m$=6 in a three-phase uncontrolled bridge rectifier, $U_d$ is the uncontrolled rectifier output voltage.

The output voltage and current of the uncontrolled rectifier can be filtered by $L_1$ and $C_1$ to obtain the input voltage and current $U_i$ and $I_i$ of the downstream DC-DC converter.

$$\begin{cases} U_i = U_d - L_1(dI_d/dt) \\ I_i = I_d - C_1(dU_i/dt) \end{cases} \tag{2}$$

The secondary voltage and current of the isolation transformer can be obtained from the actual on/off state function $D(t)$.

$$\begin{cases} U_s = \dfrac{U_\mathrm{p}}{n} = \dfrac{1}{n}\left(U_i D(t) - L_{lk} D(t)\dfrac{dI_i}{dt}\right) \\ I_s = nI_\mathrm{p} = nI_i D(t) \end{cases} \tag{3}$$

where $U_s$ and $I_s$ are the secondary voltage and current of the transformer, respectively. $L_{lk}$ is the leakage inductance.

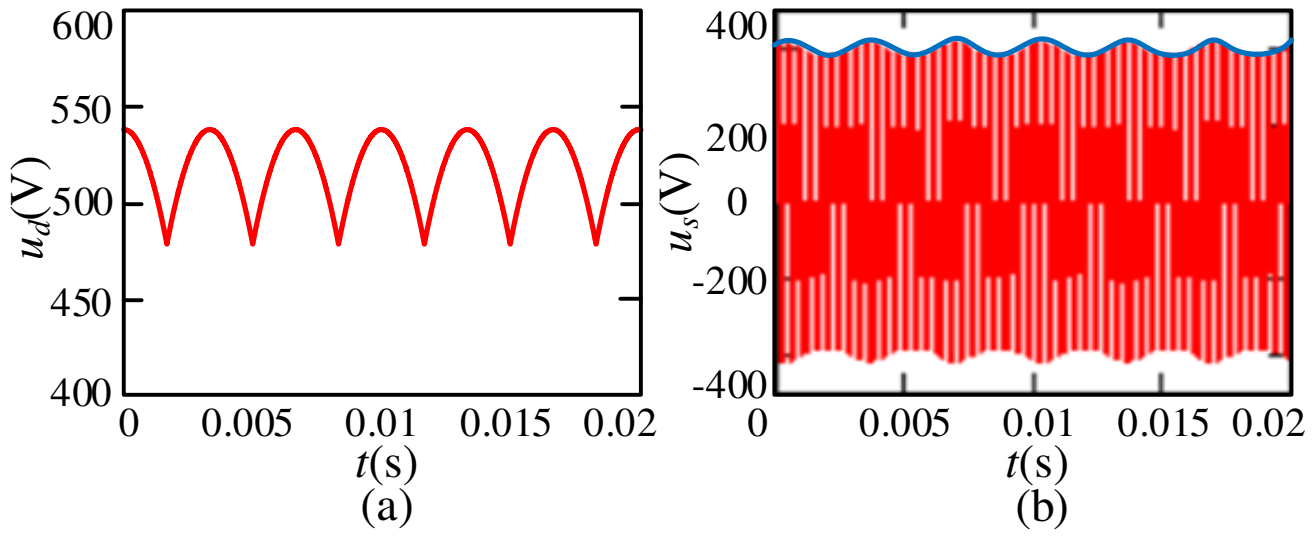

Fig. 3 Output voltage waveforms (a) Uncontrolled rectifier output voltage (b) Isolation transformer secondary voltage

Fig. 3(a) is the output voltage waveform of the uncontrolled rectifier. It can be seen from the diagram that the output voltage contains a large six-pulse component. Fig. 3(b) shows the secondary voltage waveform of the isolation transformer. It can be seen that the secondary voltage of the isolated DC-DC converter also has periodic six pulse components. Although the ripple on the transformer in the figure is not obvious, its ripple amplitude reaches 34V through calculation and simulation experiments. Then the six-pulse component causes an increase in the ripple of the output voltage. Therefore, it is of great significance to improve the steady-state ripple of the output voltage of the isolated DC-DC converter from the control aspect without adding too many filter elements.

## III. Adaptive Linear Active Disturbance Rejection Control

Linear active disturbance rejection control technology is widely used in nonlinear system control, and its control effectiveness and advantages have also been verified [10], [11]. Higher controller bandwidth is conducive to improving immunity performance. However, for the downstream DC-DC converter, the traditional LADRC control based on fixed controller bandwidth parameters cannot well solve the problem of slow output voltage response speed caused by sudden load changes. Therefore, this paper proposes an adaptive linear active disturbance rejection control method which can flexibly adjust the controller bandwidth parameters according to the real-time value of output current to solve this problem and improve the system output response speed at the same time.

### A. Control parameters design of the proposed A-LADRC

The overall control diagram of the proposed A-LADRC method is shown in Fig. 4. Based on the pole assignment method of the characteristic equation of Linear State Error Feedback control (LSEF) and Linear Extended State Observer (LESO), the parameter selection is improved considering practical engineering applications.

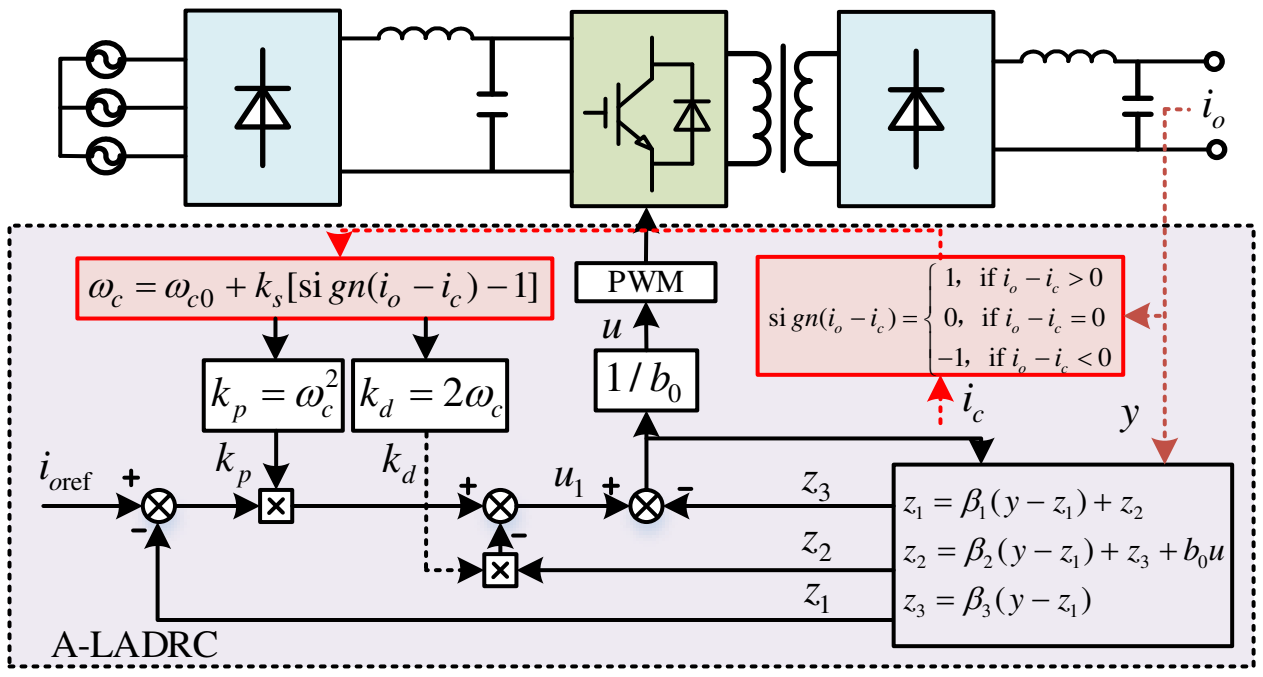

Fig. 4 The proposed A-LADRC diagram

1) Linear state error feedback control

For the second-order system studied in this paper, the LSEF is commonly used in Proportional-Derivative (PD) combination form, and the control expression can be set as

$$u_o = k_p(R - z_1) - k_d z_2 \tag{4}$$

where $u_o$ is the LSEF output controlled variable, $R$ is the input value given by the system, $z_1$ and $z_2$ are the state variables observed by LESO, $k_p$ and $k_d$ are the gains of the controller. The control rate is designed as

$$u = (u_o - z_3)/b_0 \tag{5}$$

Generally, the second-order system is considered as

$$\ddot{y} = -a_1\dot{y} - a_0 y + bu \tag{6}$$

According to the pole assignment method in [17], the controller bandwidth is $\omega_c$, and the closed-loop poles of the system can be assigned at -$\omega_c$. Then the following expression can be obtained.

$$s^2 + k_d s + k_p = (s + \omega_c)^2 \tag{7}$$

Based on (7), the controller gain can be obtained as

$$k_p = \omega_c^2, k_d = 2\omega_c \tag{8}$$

$$\omega_c = 4.75/t_s \tag{9}$$

where, $t_s$ is the system adjustment time.

Based on the traditional LADRC, the controller bandwidth $\omega_c$ is closely related to the response speed of the system, as shown in (9). The larger the $\omega_c$, the stronger the function of the controller and the faster the system response speed. However, if it is too large, it may also cause the system to overshoot and oscillate. The selection is mainly determined according to the response speed of output current, system overshoot, regulation time and waveform quality. In practice, the selection method can be used by Simulink experiment accurate range, and then by observing the output current waveform adjustment to achieve the best results. Therefore, it is easy to determine the controller bandwidth value through simulation.

Through theoretical analysis and experimental testing, the selection of system control gains $k_p$ and $k_d$ has a great impact on the system response speed. While the selection of the two parameters is related to the system bandwidth $\omega_c$. The capacity of the system load can be characterized by the load current value. Therefore, an A-LADRC method is proposed to improve the output response speed of the AC-DC-DC power supplies. It can flexibly adjust the bandwidth parameters of the controller according to the real-time value of the output current. The output current value of the load side collected in real time is compared with the set critical current value, and the bandwidth of the actual output system is flexibly controlled based on the comparison result. When the system works in the no-load state or light-load state, the load current value is smaller at this time, and the critical current value is defined as $i_c$. The designed control bandwidth adaptive adjustment expression is described as

$$\omega_c = \omega_{c0} + k_s[\text{sign}(i_o - i_c) - 1] \tag{10}$$

$$\text{si}\,gn(i_o - i_c) = \begin{cases} 1, & i_o - i_c > 0 \\ 0, & i_o - i_c = 0 \\ -1, & i_o - i_c < 0 \end{cases} \tag{11}$$

where $k_s$ is the adaptive adjustment coefficient, and $i_c$ can be set as 30.

Based on (10), when the system is in the normal operating state, the system bandwidth of the control parameter is $\omega_{c0}$, according to the traditional control parameter identification can determine its value. When the system operates in the no-load or light-load state, the optimal value of the current state can be reselected through the adjustment coefficient.

2) Linear extended state observer

LESO is the key part of the control algorithm. It has a good anti-interference effect on the nonlinear system. For the second-order system studied in this paper, the LESO expression can be expressed as

$$\begin{cases} z_1 = \beta_1(y - z_1) + z_2 \\ z_2 = \beta_2(y - z_1) + z_3 + bu \\ z_3 = \beta_3(y - z_1) \end{cases} \tag{12}$$

where $z_1$, $z_2$, $z_3$ are state variables, respectively; $\beta_1$, $\beta_2$, $\beta_3$ are observer gains. The observer can track the state variables by selecting appropriate gain parameters.

Similarly, the pole assignment method can be used to set the observer bandwidth $\omega_0$. Then the following expression can be obtained.

$$s^3 + \beta_1 s^2 + \beta_2 s + \beta_3 = (s+\omega_0)^3 \tag{13}$$

The observer gain parameter is obtained as

$$\beta_1 = 3\omega_0,\ \beta_2 = 3\omega_0^2,\ \beta_3 = \omega_0^3 \tag{14}$$

According to (8) and (14), it can be seen that the controller parameter design can be realized by adjusting the controller bandwidth $\omega_c$ and observer bandwidth $\omega_0$. The observer bandwidth $\omega_0$ is generally taken as 4-10 times of $\omega_c$.

*B. Stability analysis of the proposed A-LADRC*

According to (12), the transfer function matrix form can be obtained as

$$\begin{bmatrix} Z_1(s) \\ Z_1(s) \\ Z_1(s) \end{bmatrix} = \frac{1}{N(s)} \begin{bmatrix} b_0 s & \beta_1 s^2 + \beta_2 s + \beta_3 \\ b_0 s^2 + b_0\beta_1 s & \beta_2 s^2 + \beta_3 s \\ -b_0\beta_3 & \beta_3 s^2 \end{bmatrix} \begin{bmatrix} u \\ y \end{bmatrix} \tag{15}$$

$$N(s) = s^3 + \beta_1 s^2 + \beta_2 s + \beta_3 \tag{16}$$

Combined with (4), the transfer function of the second-order A-LADRC controller can be obtained. $G_c(s)$ is the controller, and $G_F(s)$ is the filter equivalent transfer function.

$$G_F(s) = \frac{k_p(s^3 + \beta_1 s^2 + \beta_2 s + \beta_3)}{(k_p\beta_1 + k_d\beta_2 + \beta_3)s^2 + (k_p\beta_2 + k_d\beta_3)s + k_p\beta_3} \tag{17}$$

$$G_c(s) = \frac{(k_p\beta_1 + k_d\beta_2 + \beta_3)s^2 + (k_p\beta_2 + k_d\beta_3)s + k_p\beta_3}{b_0 s[s^2 + (k_d + \beta_1)s + k_d\beta_1 + k_p + \beta_2]} \tag{18}$$

According to [24], the transformer secondary side of the phase-shifted full-bridge converter is similar to the traditional buck converter. By modeling and analyzing the small-signal equivalent model of the DC-DC converter, the transfer functions from the duty cycle of the converter to the output voltage and output current can be obtained as follows:

$$G_{uod}(s) = \left.\frac{\hat{u}_o(s)}{\hat{d}(s)}\right|_{\hat{u}_i(s)=\hat{i}_i(s)=0} = \frac{nU_i}{L_2C_2s^2 + (\frac{L_2}{R_1} + R_dC_2)s + \frac{R_d}{R_1} + 1} \tag{19}$$

$$G_{iod}(s) = \frac{G_{uod}(s)}{R} = \frac{nU_i}{RL_2C_2s^2 + R(\frac{L_2}{R_1} + R_dC_2)s + R\frac{R_d}{R_1} + R} \tag{20}$$

The controlled object transfer function $G_p(s)$ can be obtained from (20), so the open-loop transfer function of the system can be deduced as

$$G_k(s) = G_F(s)G_c(s)G_p(s) \tag{21}$$

The closed-loop transfer function of the system is as follows:

$$G_b(s) = \frac{G_F(s)G_c(s)G_p(s)}{1 + G_c(s)G_p(s)} \tag{22}$$

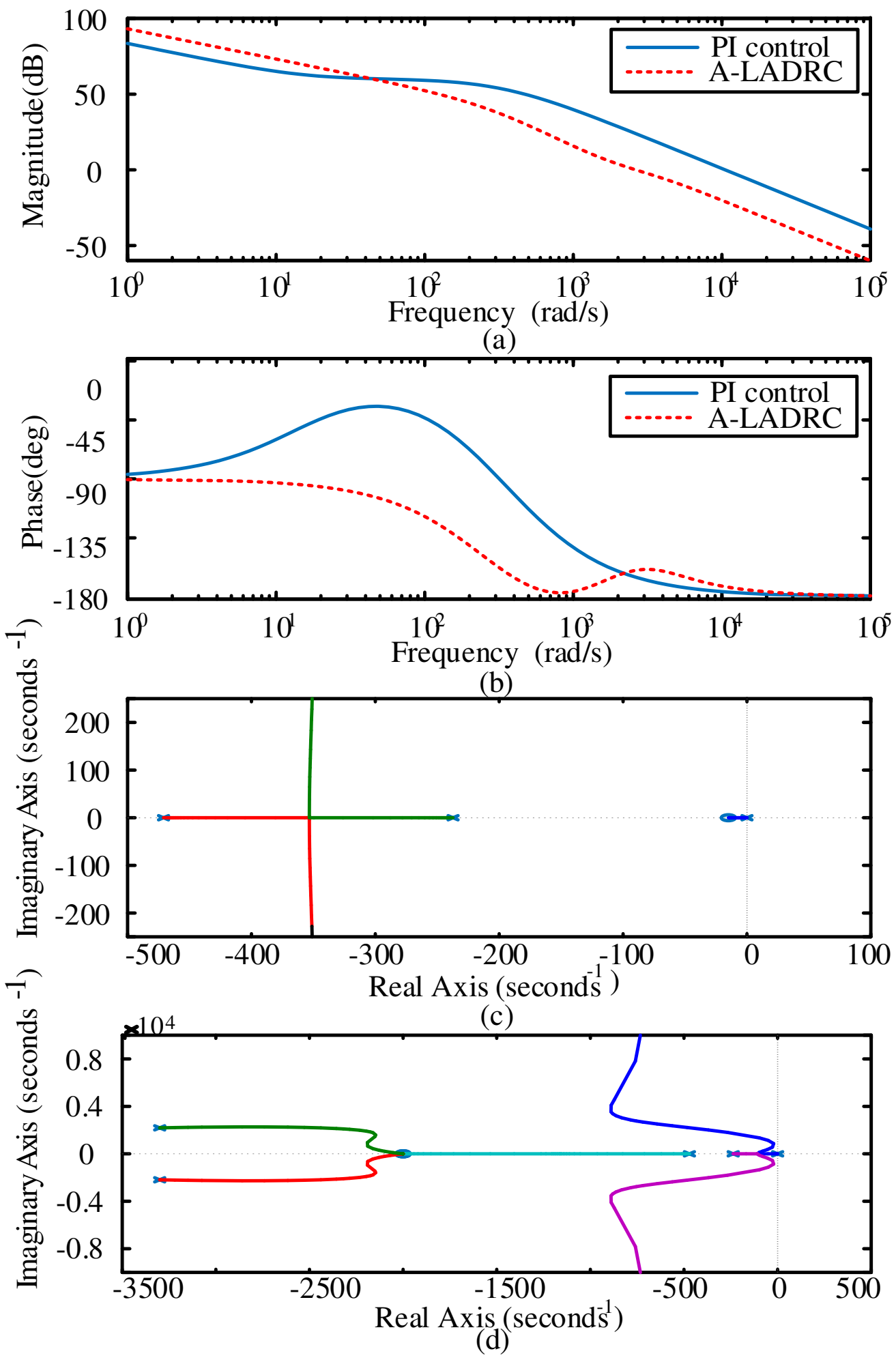

Fig. 5 System stability analysis (a) Amplitude-frequency characteristics (b) Phase-frequency characteristics (c) PI control root locus (d) A-LADRC root locus

In order to verify the control performance of the proposed A-LADRC controller, the open-loop transfer function bode diagram and root locus diagram of the system under the A-LADRC and the PI control are shown in Fig. 5. The frequency domain characteristics of the two methods are compared. It can be seen from the diagram that in the low-frequency band, the amplitude of A-LADRC curve is larger, the command tracking effect is stronger, and the low-frequency disturbance suppression ability is better. In the high-frequency band, the amplitude of A-LADRC curve is lower, and the suppression ability of high-frequency noise is better than PI control.

## IV. Output Voltage Ripple Reduction Control

When the AC grid supplies power to the DC load, the DC bus voltage obtained by the uncontrolled rectifier has six-pulse components, resulting in the same six-pulse fluctuation components of the downstream output voltage. The output voltage ripple can also be reduced by adding additional capacitors and inductors [25], [26]. However, considering the cost and the implementation method, this paper designs a duty cycle compensation control strategy for a single three-phase isolated AC-DC-DC module. For a twelve-module IPOP system, a Hierarchical Delay Current Sharing Control (HDCSC) strategy of the centralized controller is proposed. The specific analysis is as follows.

### *A. Duty cycle compensation control for a module*

The power supply designed in this paper is mainly used in resistive load scenarios, and the magnitude of the voltage ripple can also be reflected by the current ripple. Therefore, based on the A-LADRC strategy for a single AC-DC-DC module, the compensation amount of the converter duty cycle can be given according to the detected output current deviation, and the A-LADRC can be assisted to fine-tune the system output voltage. The duty cycle compensation $D_c$ can be designed as:

$$D_c = k_w \frac{i_{\mathrm{ref}} - i_o}{i_{\mathrm{ref}}} \tag{23}$$

$$D = D_L + D_c \tag{24}$$

where $i_{\mathrm{ref}}$ is the given output reference current of the system and $k_w$ is the compensation coefficient. Too large kW selection will cause excessive compensation and aggravate oscillation, but too small can not achieve compensation effect. The selection is mainly determined according to the size of the current set value, the fluctuation degree of the system voltage and the compensation target value. In order to avoid excessive compensation, a limiting link is added in the control loop. Finally, the value can be determined as 0.01 by simulation.

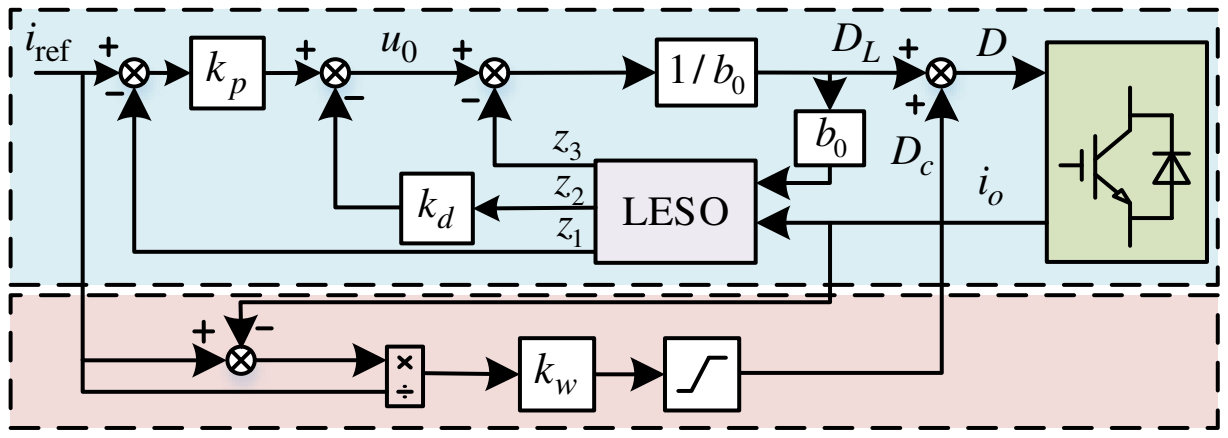

Fig. 6 Duty cycle compensation control block diagram

The duty cycle compensation control block diagram is shown in Fig. 6. The general principles of the design method are as follows. When the system output current is higher than the reference current, the upward fluctuation trend of the output current is suppressed by slightly reducing the duty cycle. Similarly, when the output current is lower than the reference current, the decreasing trend of the output current is suppressed by slightly increasing the duty cycle.

Comparing waveforms with and without duty cycle compensation are shown in Fig. 7. It can be seen that the real-time control compensation through the deviation between the output current and the reference value can be achieved with the designed duty cycle compensation control strategy. The larger the deviation is, the greater the compensation value is. And the compensation coefficient can be adjusted to control the compensation value. From the comparison waveforms of the duty cycle with and without compensation in actual operation, it can be seen that when the current is at the peak of ripple, the switching tube conduction time is shortened. When it is at the trough of ripple, the switching tube conduction time is extended, thus achieving a better compensation effect. By selecting the appropriate compensation coefficient, the output fluctuation of the system can be suppressed to the maximum extent, reducing the output ripple. It can also speed up the regulation of DC voltage and current, reduce the voltage sag or rise amplitude, and finally realize the stability of output voltage and current.

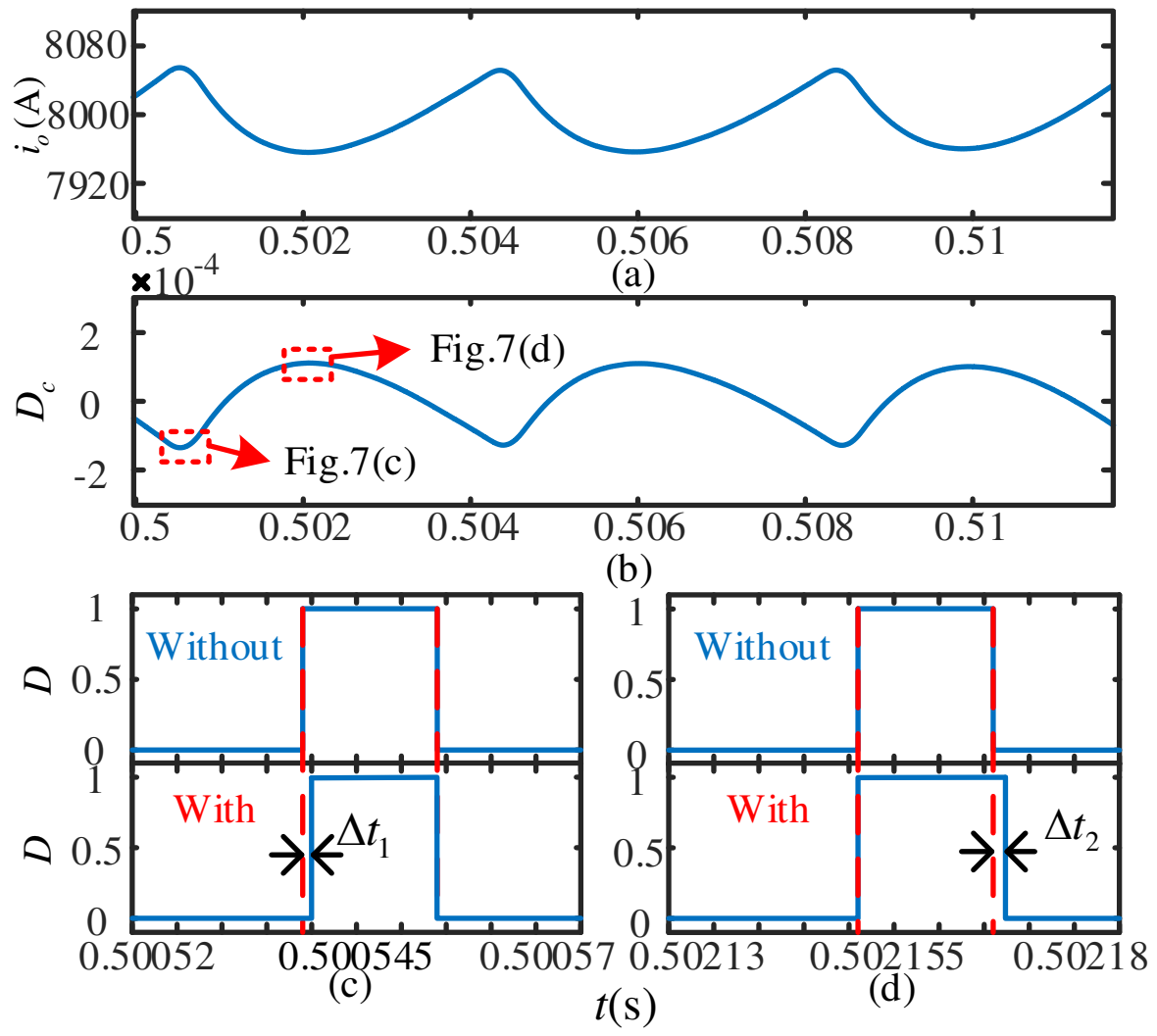

Fig. 7 Comparison of waveforms with and without duty cycle compensation (a) Output current waveform (b) Duty cycle compensation value (c) Amplified waveforms with and without duty cycle compensation (d) Amplified waveforms with and without duty cycle compensation

*B. Hierarchical delay current sharing control for multiple IPOP modules*

For the designed twelve-module IPOP system in this paper, the DC supply with a parallel structure can form a high-power supply and increase the operation stability of the system. For the multi-module power parallel system, this paper uses 42V/12000A prototype platform to test the performance of multi-group current sharing under the traditional PI control and A-LADRC control. The experimental results show that the system output current can be equalized through the combined delay control of each module, which can effectively reduce the output ripple. Therefore, in this paper, the twelve modules are divided into four groups, and the unit controllers of the four groups are adopted. Three modules are controlled by a unit controller. Based on the above improved A-LADRC and duty cycle compensation, the HDCSC strategy is adopted to control the delay of the module and the unit controller at different times, so as to reduce the output ripple of the system in two levels.

Fig. 8(a) shows the control schematic diagram of the twelve modules IPOP system. Taking a single unit controller as an example, the control schematic block diagram of a single unit controller is shown in Fig. 8(b). One unit controller controls three modules and makes different control delays $t_1$, $t_2$ and $t_3$ for the three modules respectively. Similarly, the time delay $\tau_1$, $\tau_2$, $\tau_3$ and $\tau_4$ among the four-unit controllers are set. Reasonable selection of the delay time can make the peak values of each unit's current ripple interleaved, so as to achieve the purpose of reducing the system overall output ripple.

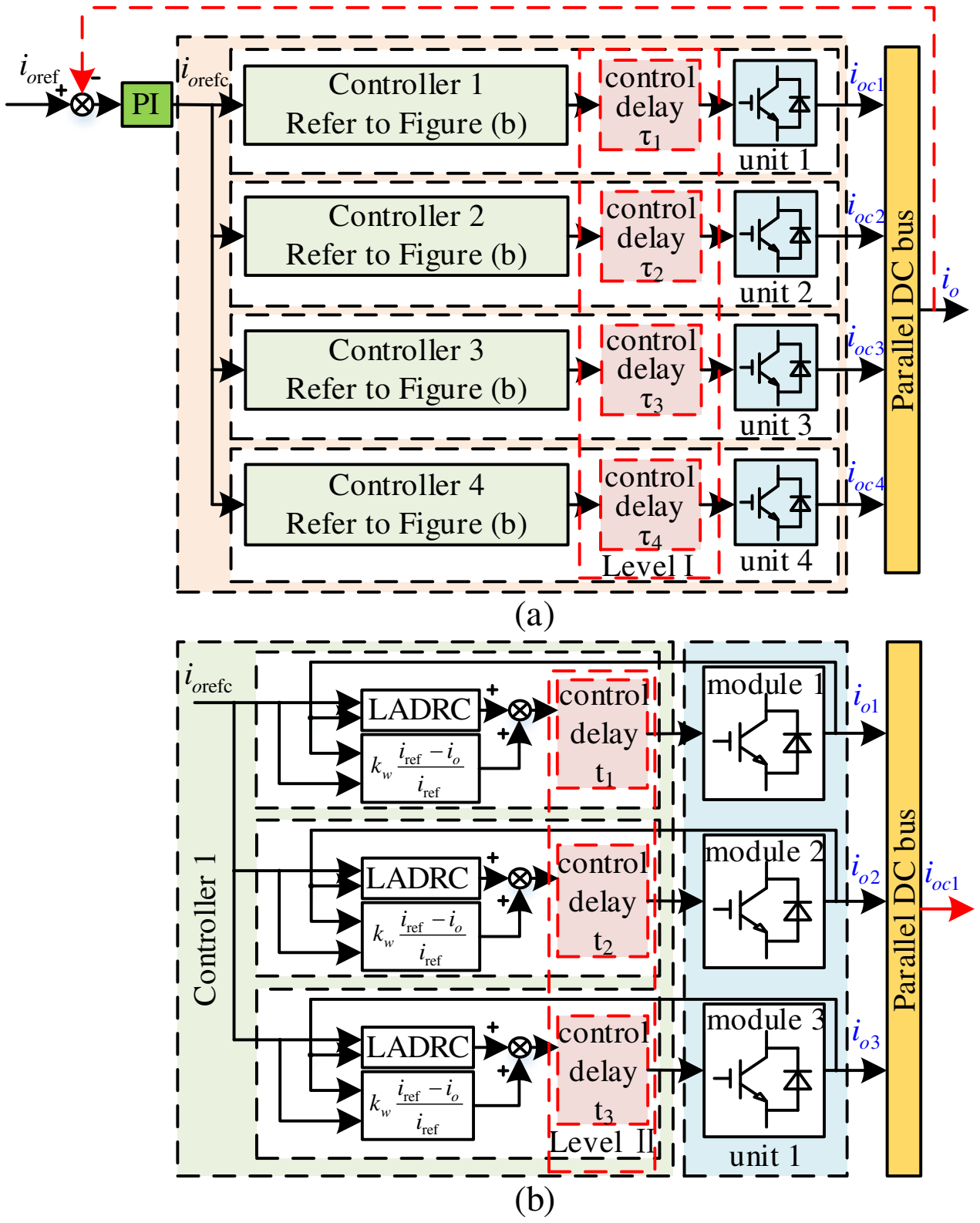

Fig. 8 Schematic diagram of the system hierarchical delay current sharing control (a) control schematic diagram of the overall parallel system (b) a single controller control schematic diagram

According to the previous calculation, due to the influence of the front-end uncontrolled rectifier, the output DC voltage has voltage fluctuation all the time. And the variation period of one pulse component is $T_m$=3.3ms. When there are $x$ modules in each group of controllers with first-stage delay, it can be calculated from (25) that the first-stage delay period td1=1.1ms. For the three modules controlled by a single controller, the controller output ripple is minimum when $t_1$=0ms, $t_2$=1.1ms and $t_3$=2.2ms. The effect of delay control can be referred to Fig. 9.

$$t_{d1} = \frac{T_m}{x} \tag{25}$$

$$t_{d2} = \frac{t_{d1}}{y} \tag{26}$$

By the same token, the delay time of the second-stage controller can be selected. After the flow sharing control among the first-stage modules, ripple pulsation period is about 1.1ms. In the case of controller $y$ group, $t_{d2}$=0.275ms can be obtained from (26), so $\tau_1$=0ms, $\tau_2$=0.275ms, $\tau_3$=0.55ms and $\tau_4$=0.825ms can be selected.

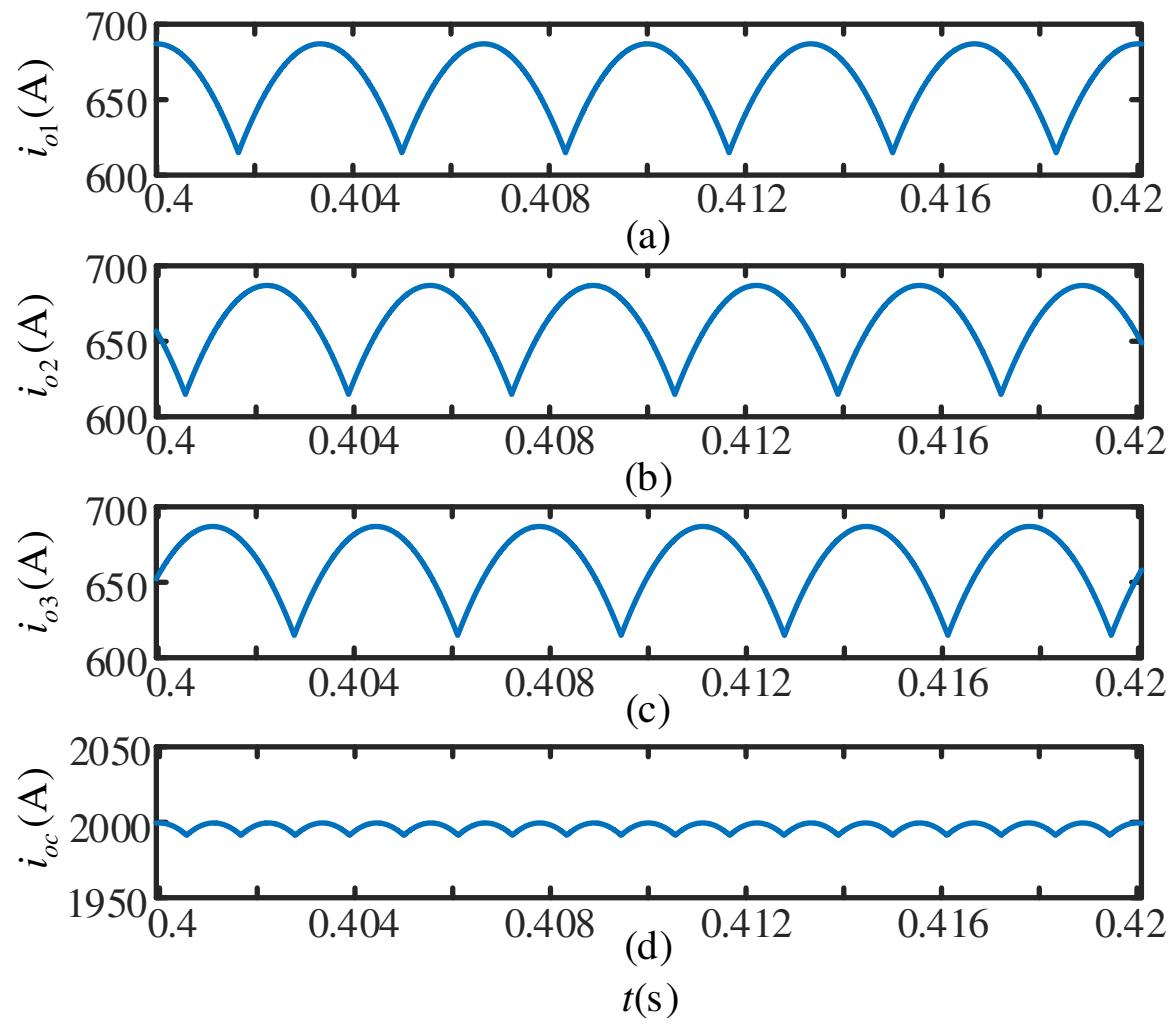

Fig. 9 Delay control output waveforms (a) Module 1 output current (b) Module 2 output current (c) Module 3 output current (d) Controller 1 output current

## V. Experimental Results

### A. Hardware-in-the-loop experimental results

In order to verify the effectiveness of the proposed A-LADRC method and the duty cycle compensation strategy, in this paper, a Hardware-In-the-Loop (HIL) platform is used to verify the results which are difficult to obtain with an experimental prototype, as shown in Fig. 10. The HIL platform is mainly composed of StarSim MT6020 simulator and RTU-BOX real-time digital controller. The advantages of the proposed A-LADRC control method are mainly reflected by the comparison with the commonly used PI control strategy. Considering that this paper focuses on A-LADRC and ripple compensation control technology, the system response time and waveform quality are compared with PI control and LADRC control in the experiment under the same response speed, so the bandwidth of PI controller is not analyzed in detail. The main system and control parameters are shown in Tab. 1.

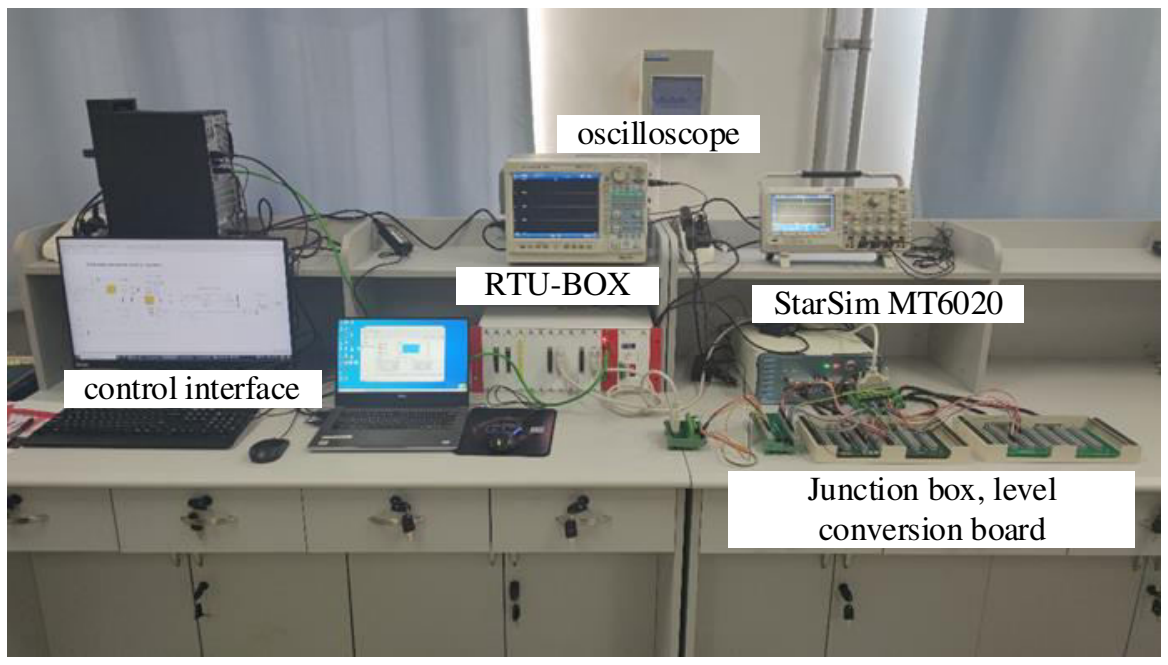

Fig. 10 Hardware-in-the-loop experimental platform

Table 1 System and control parameters

| Parameter | Value | Parameter | Value | Parameter | Value |
|---|---|---|---|---|---|
| $u_{abc}$ | 380V | $L_1$ | 3200μH | $R_1$ | 0.01Ω |
| *ratio* | 1:1.5 | $C_1$ | 3000μF | $\omega_c$ | 400 |
| $f_s$ | 15kHz | $L_2$ | 3000μH | $\omega_0$ | 2800 |
| $L_{lk}$ | 30μH | $C_2$ | 3500μF | $k_w$ | 0.01 |

Under the condition that the system and typical parameters are consistent, the PI control, the LADRC and the proposed A-LADRC with ripple reduction control are adopted to compare the output results, respectively. Fig. 11(a) is the experimental waveform at the start time. It can be found that all three control methods can realize the tracking control of a given output current reference of 8000A. However, the system output under A-LADRC control takes the shortest time to reach a stable state, which is about 0.03s. While the time for the system under PI control to reach a stable state is about 0.15s. In addition, the response speed of the proposed A-LADRC is the fastest. At the same time, in the steady state, the added ripple suppression strategy can well reduce the fluctuation of the system.

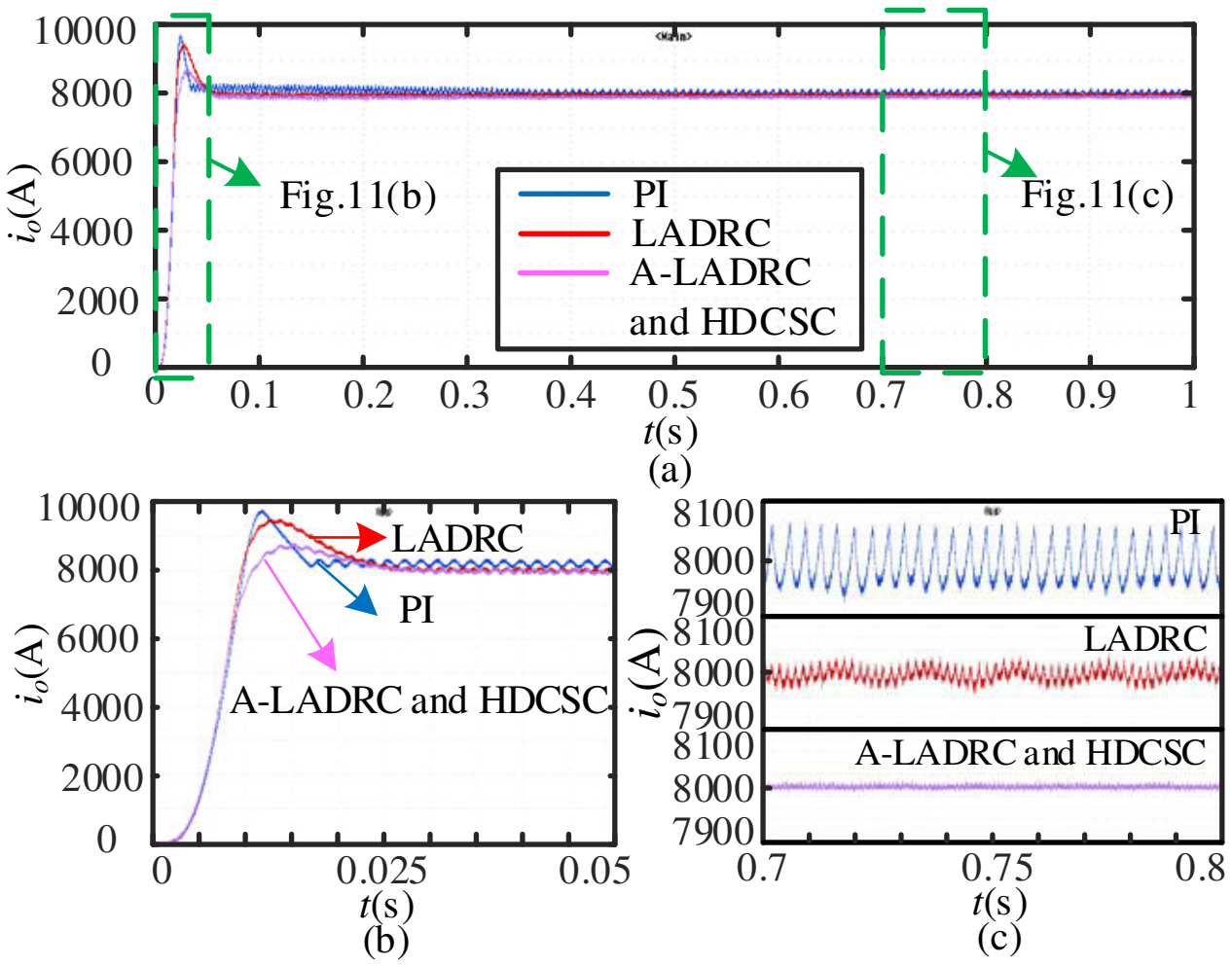

Fig. 11 Output current waveforms under different control methods (a) Full response process (b) Experimental waveforms at startup time (c) Experimental waveforms at steady state

As shown in Fig. 11(c), the diagram shows the steady-state output current amplified waveforms of the system under different controls. It can be seen that when the system adopts PI control, the amplitude of the output current ripple is about 128A. When the LADRC strategy is adopted, the output current not only has better following characteristics, but also the output current ripple fluctuation amplitude is reduced to 60A, which is 50% lower than that of PI control. Finally, after adopting the proposed A-LADRC control and adding the designed ripple improvement strategy, it can be seen that the output six-pulse component is significantly reduced. The output current ripple fluctuation amplitude is about 12A, which is reduced by 90% and 80% compared to PI control and LADRC control, respectively. This verifies the significant superiority of the ripple improvement strategy proposed in this paper under high-power operating conditions.

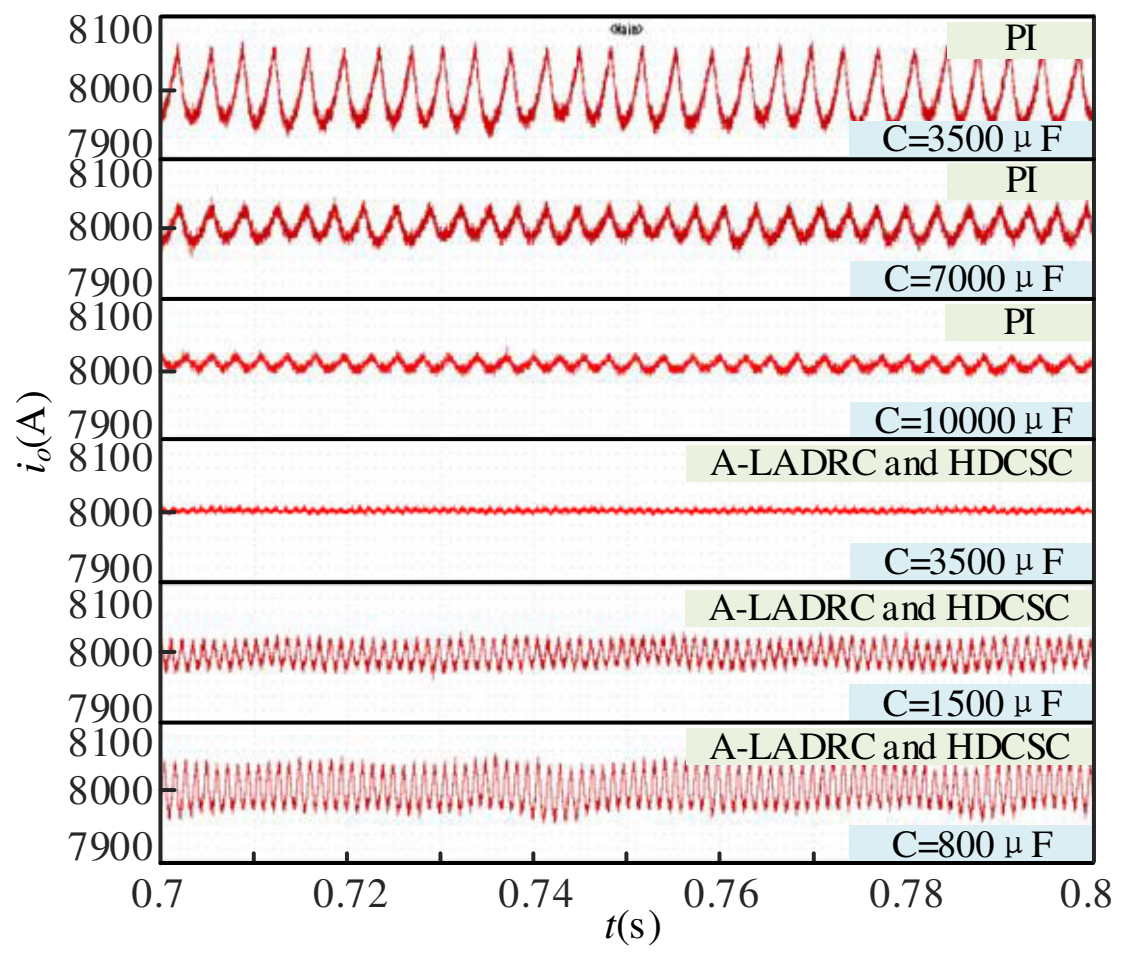

Fig. 12 Output current waveforms with different filter capacitors and control strategies.

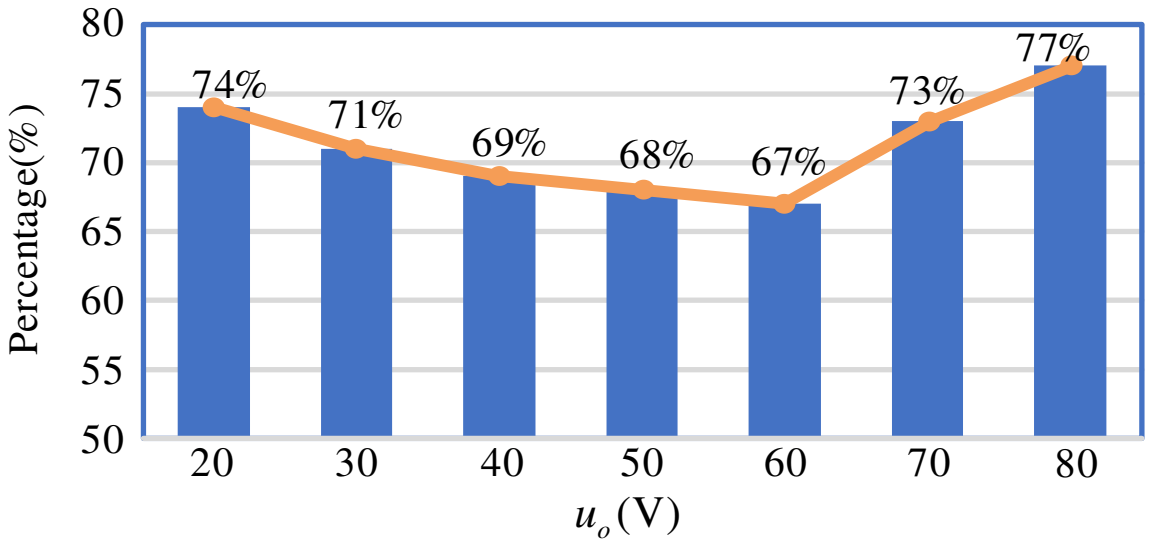

Fig. 13 Reduced ratio of the output filter capacitor with the proposed control when the ripple amplitude is within 1.6%.

To further verify the superiority of the ripple reduction control designed in this paper, the output filter capacitor $C_2$ is changed under different control modes. The output current waveforms under the PI control and the proposed A-LADRC with HDCSC are shown in Fig. 12. The initial designed value of the filter capacitor $C_2$ is 3500μF. At this time, when the system adopts PI control, the output current ripple fluctuation amplitude is about 128A. When A-LADRC control is adopted, the current ripple fluctuation amplitude is only about 12A, and the ripple amplitude is reduced by about 90%. As can be seen from Fig. 12, the output ripple gradually decreases with the increase of the filter capacitor under the PI control. However, with the increase of the capacitor value, the ripple reduction effect is significantly reduced. While the application cost is greatly increased at the same time. On the contrary, under the proposed control of the A-LADRC with HDCSC, a good ripple suppression effect can still be achieved with a small filter capacitor. The ripple value under the designed control strategy with an 800μF filter capacitor is basically the same as the ripple caused by PI control with a 3500μF filter capacitor. Therefore, the proposed strategy can effectively reduce the cost of the filter capacitor. If the ripple fluctuation range is set within 1.6% for different output voltages, to obtain the reduced capacitance value by the designed strategy, the capacitance reduction percentage values are given through multiple tests on the HIL platform, as shown in Fig. 13. It can be seen that the control strategy proposed in this paper can maintain the stable output of the system with a small ripple, which greatly reduces the filter capacitor in different voltage levels. Thus, the effectiveness of the ripple improvement strategy proposed in this paper is further verified.

*B. 42V/12000A prototype experimental results*

In order to verify the effectiveness and superiority of the proposed control method, a 42V/12000A prototype with twelve IPOP modules is designed, which is shown in Fig. 14.

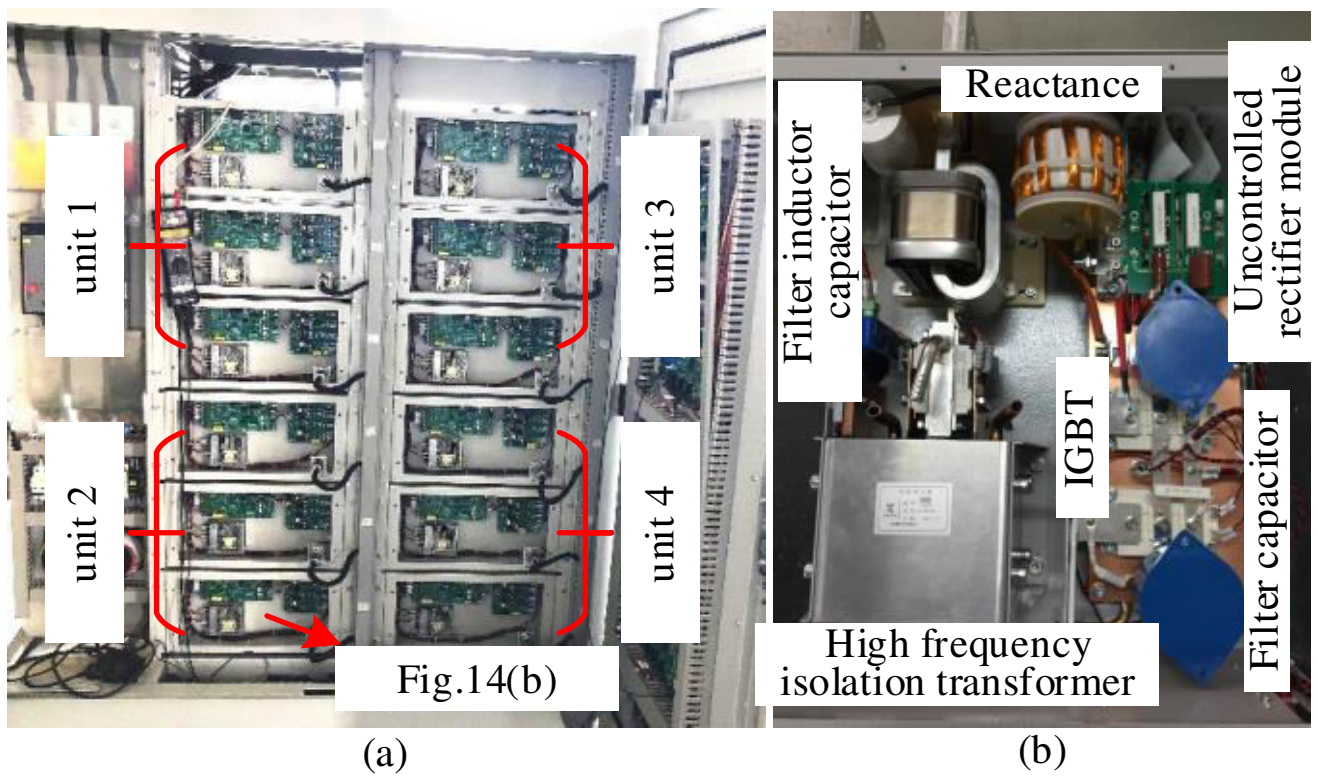

Fig. 14 A 42V/12000A prototype (a) Diagram of the overall system (b) Internal structure of a module

The prototype experimental test conditions are set as follows. At the initial state, the output voltage is controlled at 7V, and the load resistance $R_1$ is 10kΩ. At this time, the system can be equivalent to a no-load state. At about 4.8s, the load resistance decreases from 10kΩ to 3.5mΩ. Then the system switches from no-load to heavy-load mode.

Fig. 15 shows the experimental waveforms when the load is suddenly changed, and a total of 10s data is recorded. $u_{abc}$ and $i_{abc}$ are the input voltage and current of the three-phase bridge uncontrolled rectifier respectively. $u_i$ is the input voltage of the isolated DC-DC converter. $u_o$ and $i_o$ are the output voltage and output current of the isolated DC-DC converter respectively. The whole dynamic process of the system response is shown in Fig. 15(a). In order to better observe and compare the response effects after the load change, the instantaneous and steady-state waveforms of the load change are enlarged, as shown in Fig. 15(b) and Fig. 15(c), respectively. It can be seen that at the moment when the load suddenly changes, there is a peak in the output current due to the decrease in the load resistance value of $R_1$, and the output current increases from 0.05A to 2000A. The output voltage has a serious sag. The sag range reaches 80%, that is, it drops from 7V to 1.4V. And the sag lasts for more than 1s, which is caused by the slow regulation speed of the PI control. When the system reaches a new steady state, due to the limited voltage stabilization effects of the filter capacitor, the ripple amplitude of the DC bus voltage $u_i$ increases, and the peak-to-peak value of the ripple reaches 100V. Finally, this also results in a load voltage with the same large six-pulse fluctuation component, and the ripple amplitude is around 1.48V.

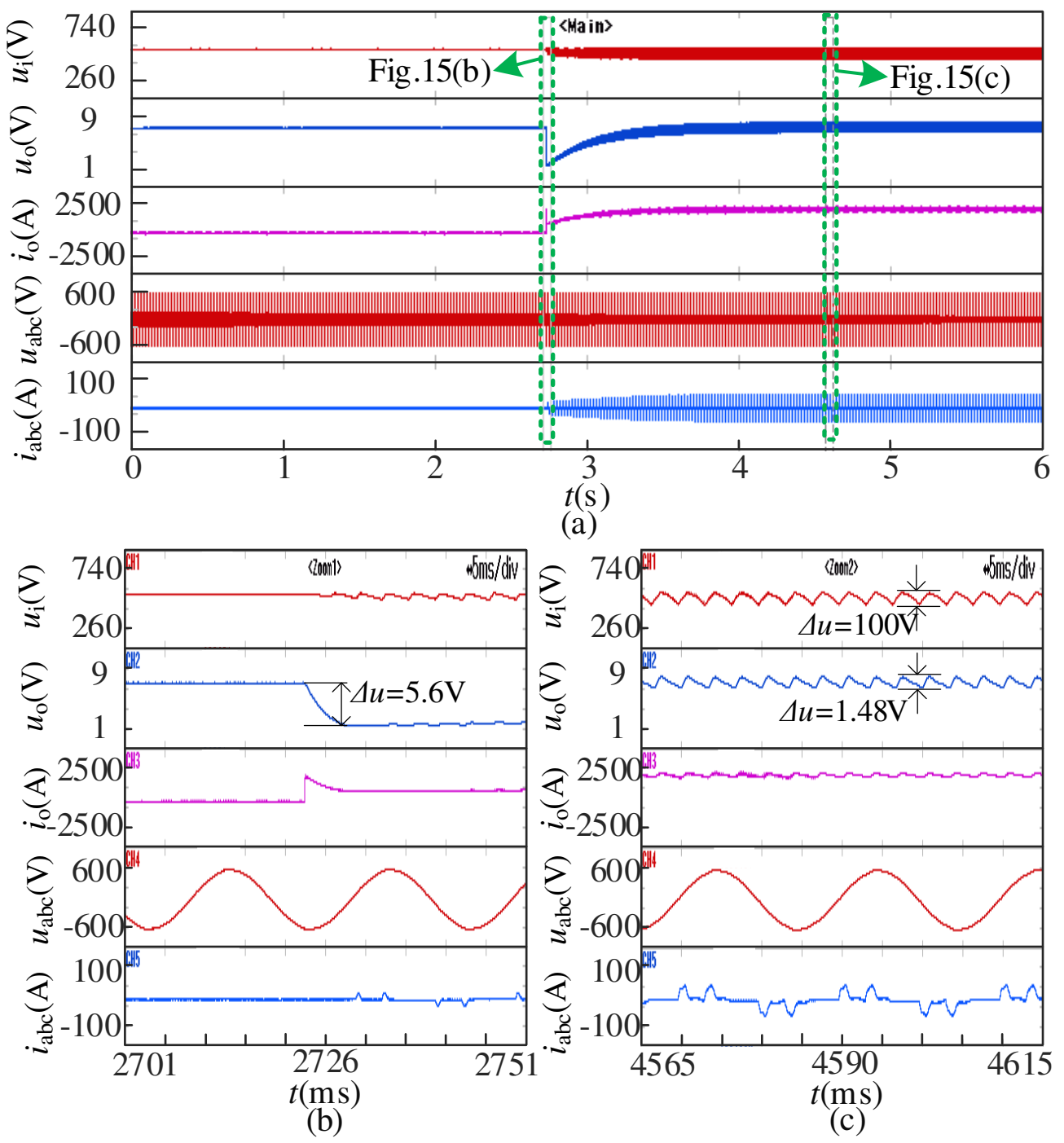

Fig. 15 Experimental waveform under PI control (a) Full response process (b) Amplified waveforms when sudden load change (c) Amplified waveforms at steady state

Fig. 16 shows the experimental waveforms under the proposed A-LADRC control, and the operating conditions are the same as those in Fig. 15.

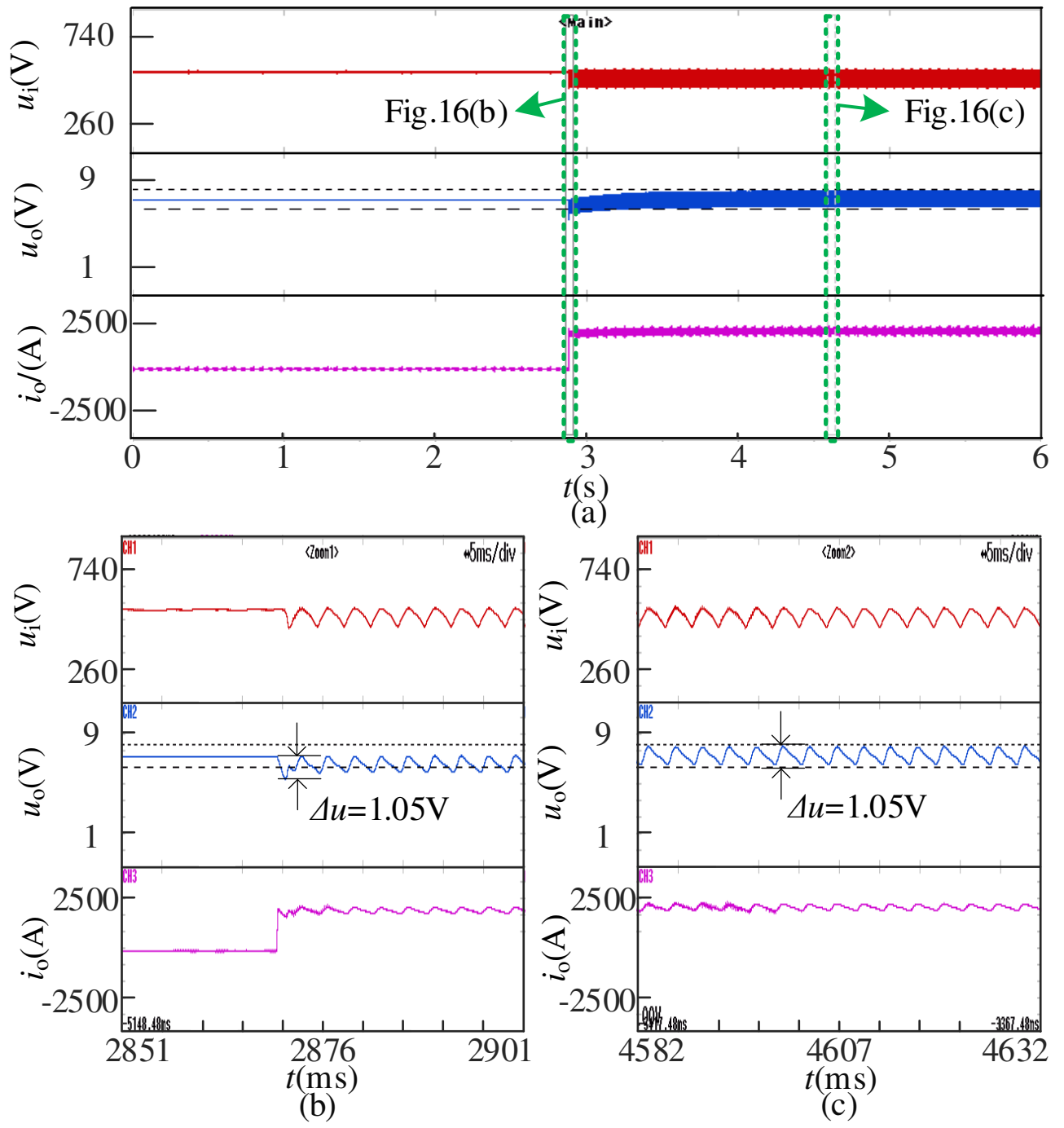

Fig. 16 Experimental waveforms under the A-LADRC control (a) Full response process (b) Amplified waveforms when sudden load change (c) Amplified waveforms at steady state

It can be seen that at the moment of load changes, the output voltage changes suddenly, and the voltage drop amplitude is about 1.05V. Compared to a drop of 80% under the PI control, the drop range is about 15%, which significantly reduces the voltage drop amplitude. At the same time, the sag duration is only 1~2ms, which greatly shortens the required time to adjust the output voltage and greatly improves the voltage response speed. It can be verified that the proposed A-LADRC strategy can effectively improve the voltage mutation when the load changes and improve the system response speed. However, the large ripple of the load voltage has not been well reduced.

In order to further reduce the ripple of load voltage, the proposed voltage ripple reduction control is added for the experiment test. The operating conditions are the same as those in Fig. 15, and the experimental waveforms are shown in Fig. 17. After the HDCSC strategy is adopted, the ripple voltage decreased from 1.48V under PI control and 1.05V under A-LADRC control to 0.31V, which decreased by 79% and 70%, respectively. It can be seen that the voltage ripple reduction control proposed in this paper has a significant effect in reducing the ripple amplitude and improving voltage quality.

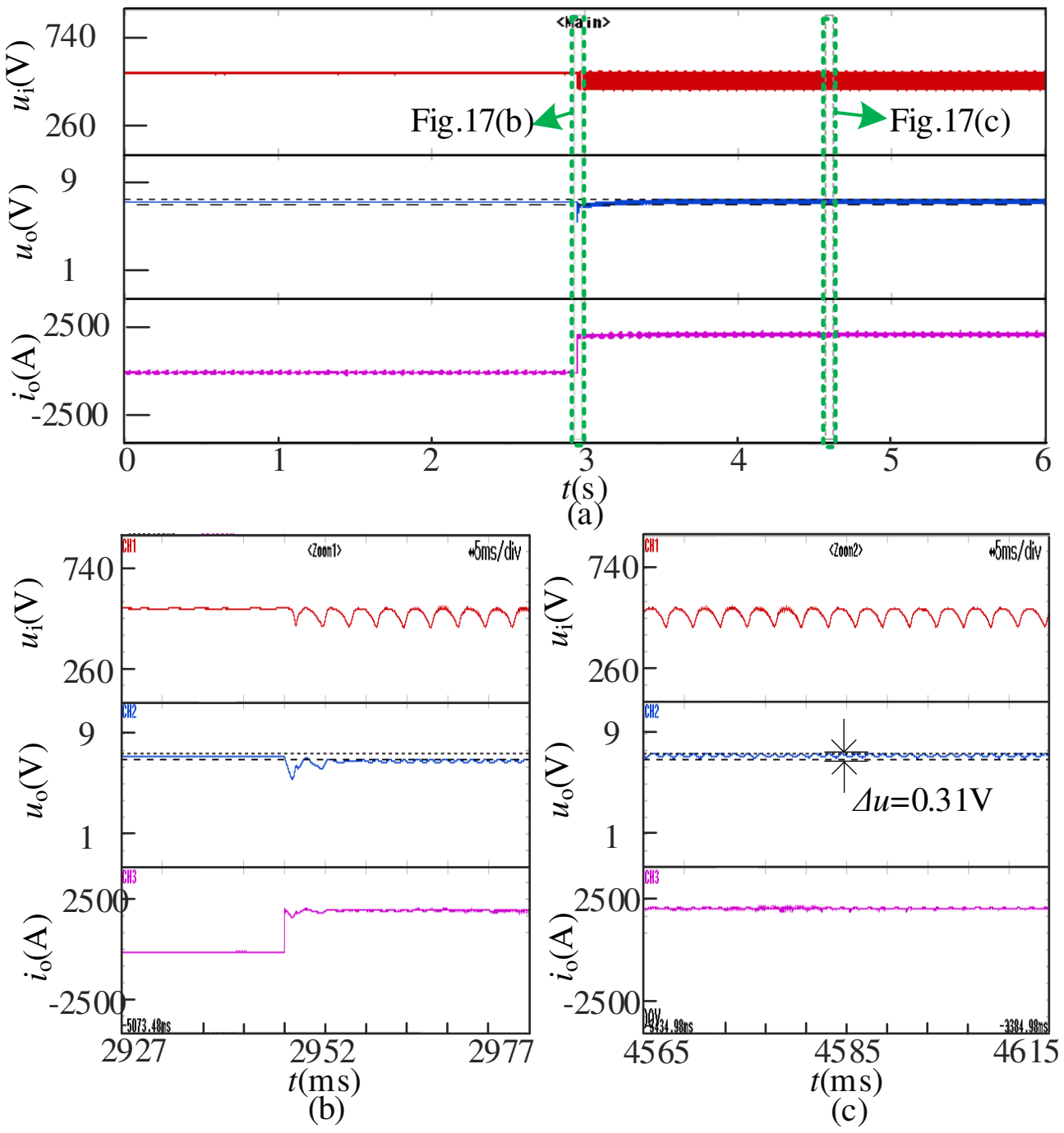

Fig. 17 Experimental waveforms with the ripple improvement method (a) Full response process (b) Amplified waveforms when sudden load change (c) Amplified waveforms at steady state

## VI. CONCLUSION

The three-phase isolated AC-DC-DC power supply is widely used in industry. To improve response speed and reduce ripple in the application of twelve IPOP modules, the A-LADRC and two ripple reduction control strategies are proposed. The following conclusions can be drawn.

1) It can be concluded from the experimental data that the A-LADRC method proposed in this paper can flexibly adjust

the bandwidth parameters of the controller according to the real-time value of the output current. Compared with the PI control, the time to achieve stability is reduced by about 80%, the overshoot is significantly reduced, and the response speed is greatly improved.

2) In order to reduce the output voltage ripple of the system, a duty cycle compensation control strategy is designed for a single-module system. For the twelve-module parallel system, a hierarchical delay current sharing control strategy is proposed. The experimental results show that the ripple improvement strategy proposed in this paper can effectively reduce the ripple amplitude by about 90% compared with PI control. While ensuring a perfect output waveform, the voltage stabilizing capacitance can be reduced by 70%-80%.

3) The performance of the proposed control is fully verified by the controller hardware-in-the-loop platform and 42V/12000A IPOP high-power DC supply, which has high guiding value in practical industrial applications.

In the application of a high-power power supply containing multiple IPOP modules, this paper proposes innovative control methods to improve the system response speed and reduce ripple from the level of control methods. Although the proposed control method has shown promising results, selecting multiple parameter values can be a complex process. To address this challenge, many studies have focused on reducing output ripple by improving circuit topology. Future research could aim to further enhance the output response time and reduce output ripple by combining the control strategy with an improved topology.

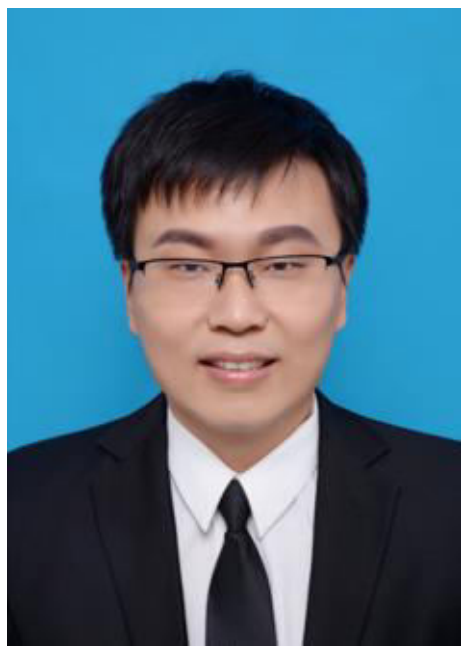

Jianhui Meng(Member, IEEE) received the B.Sc. degree in electrical engineering and automation from the North University of China, Taiyuan, China, in 2008, and the M.Sc. degree in power electronics and power drives from Dalian Jiaotong University, Dalian, China, in 2011, and the Ph.D. degree in electrical engineering from North China Electric Power University, Beijing, China, in 2015. He is currently an Engineer with the School of Electrical and Electronic Engineering, North China Electric Power University. His research interests include inverter control for distributed generator and the application of power electronic technology in power system.

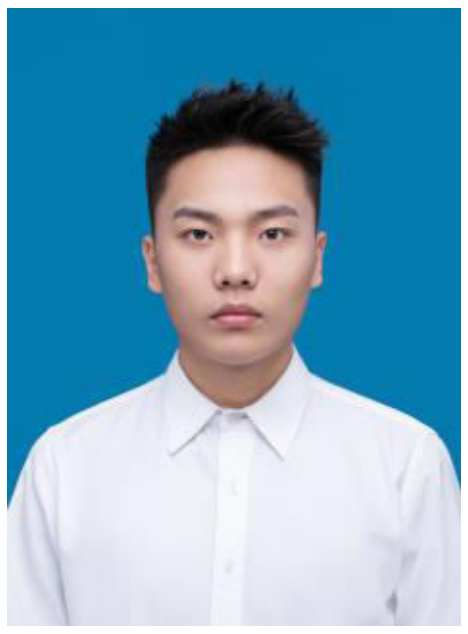

Xiaolong Wu was born in Anhui Province, China, in 1997. He received the B.S. degree from China University of Mining and Technology, Xuzhou, China, in 2020. He is currently working toward the M.S. degree in electrical engineering with North China Electric Power University, Baoding, China.
His research interests include frequency active and rapid support technology of new energy power generation system and power electronics high power supply control technology.

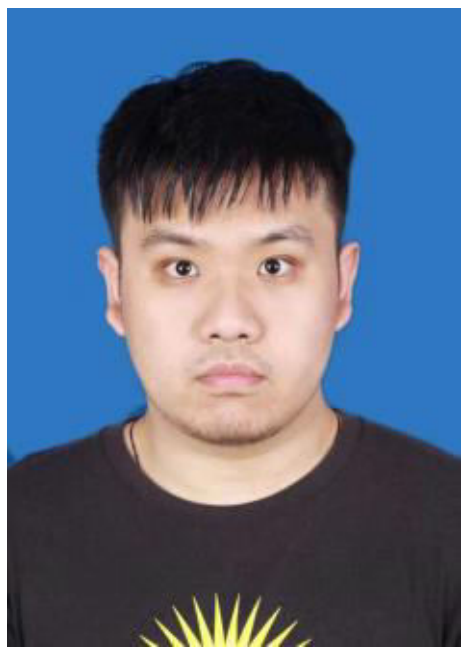

Tairan Ye received the bachelor's degree from the College of Electrical Engineering,
Nanjing Normal University, Nanjing, China in June 2021.From June 2021 to 2023, he studied for master's degree from North China Electric Power University, Hebei, China. His research interests include stability of renewable energy generators

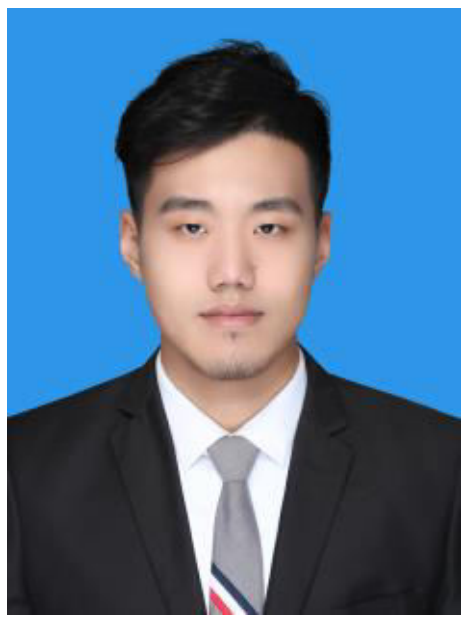

Jingsen Yu was born in Hebei Province, China, in 1997. He received the B.S. degree from Shanxi Agricultural University of Jinzhong, China, in 2020. He is currently working toward the M.S. degree in electrical engineering with North China Electric Power University, Baoding, China.
His research interests include frequency active and rapid support technology of new energy power generation system and power electronics high power supply control technology.

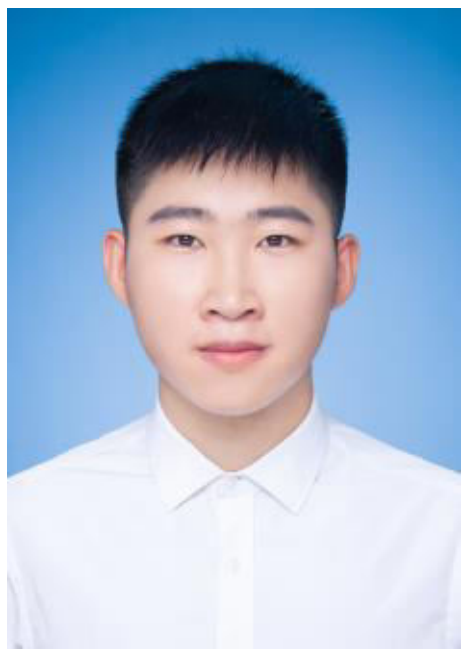

Likang Gu was born in Shandong Province, China, in 1997.He received the B.S. degree from East China Jiao Tong University, China, in 2020.He is currently working toward the M.S. degree in electrical engineering with North China Electric Power University, Baoding, China. His research interests include virtual inertial control and coordinated control of hybrid energy storage in DC networks.

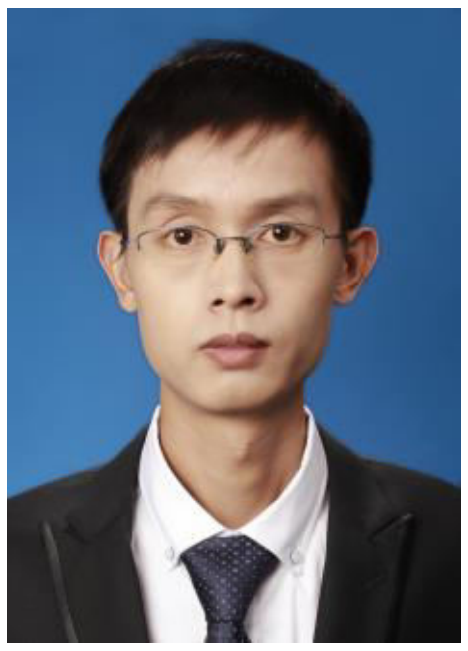

Zili Zhang received the Ph.D. degree from the College of Electrical and Electronic Engineering, North China Electric Power University, Beijing, China, in June 2015. From July 2015, he is currently a Senior Electrical Engineer of State Grid Handan Electric Power Supply Company. His research interests include grid-integration of large-scale renewable energy, and planning of electric power system.

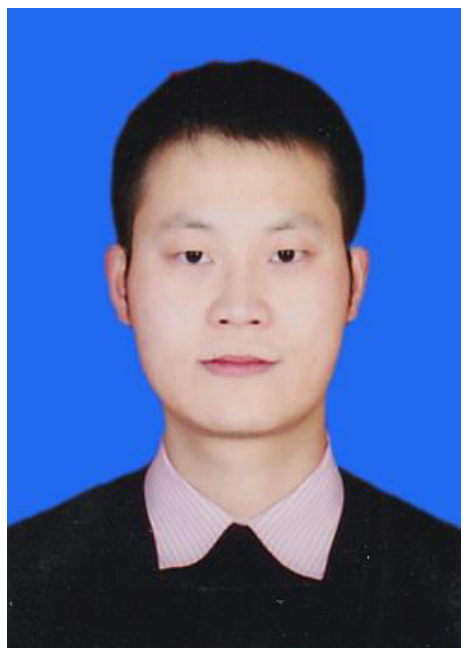

Yang Li (S'13–M'14–SM'18) was born in Nanyang, China. He received his Ph.D. degree in Electrical Engineering from North China Electric Power University (NCEPU), Beijing, China, in 2014.
He is a professor at the School of Electrical Engineering, Northeast Electric Power University, Jilin, China. From Jan. 2017 to Feb. 2019, he was also a postdoc with Argonne National Laboratory, Lemont, United States. His research interests include renewable energy integration, AI-driven power system stability analysis, and integrated energy system. He is featured in Stanford University's List of the World's Top 2% Scientists for the year 2022. He serves as an Associate Editor for the journals of IEEE Transactions on Industry Applications, and IET Renewable Power Generation. He is also a Young Editorial Board Member of Applied Energy.